\documentclass{PoS}
\pdfoutput=1
\usepackage{amsmath}
\title{Heavy quark scaling of $B\to\pi\ell\nu$ form factors with M\"{o}bius domain wall fermions}

\ShortTitle{Heavy quark scaling of $B\to\pi\ell\nu$ form factors with M\"{o}bius domain wall fermions}

\author{\speaker{Brian Colquhoun}$^a$, Shoji Hashimoto$^{a,b}$, Takashi Kaneko$^{a,b}$\\
  \llap{$^a$}High Energy Accelerator Research Organization (KEK),\\
  Tsukuba 305-0801, Japan\\
  \llap{$^b$}School of High Energy Accelerator Science, SOKENDAI (The Graduate University for Advanced Studies),\\
  Tsukuba 305-0801, Japan\\
  E-mail: \email{brian.colquhoun@kek.jp}}

\abstract{We report on the progress of our calculation of form factors
for the exclusive semileptonic decay of $B$ mesons to pions on
$2+1$ flavour lattices with spacings from $0.080~\mathrm{fm}$ down to $0.044~\mathrm{fm}$. 
Using the M\"{o}bius domain wall fermion action for all quarks, we simulate
pions with masses down to $230~\mathrm{MeV}$ and extrapolate to the physical
bottom quark mass by utilizing a range of heavy quark masses
up to $2.44$ times the mass of the charm quark. We discuss the
dependence on the pion mass, heavy quark mass and lattice spacing
in our form factors results.}

\FullConference{The 36th Annual International Symposium on Lattice Field Theory - LATTICE2018\\
		22-28 July, 2018\\
		Michigan State University, East Lansing, Michigan, USA.}

\begin{document}

\section{Introduction}
An important process for the extraction of the $|V_{ub}|$ element from the Cabibbo-Kobayashi-Maskawa matrix is $B\to\pi\ell\nu$. We report on our study of this decay, which forms part of a larger series of studies of heavy quark processes, including $B\to D^{(*)}\ell\nu$~\cite{Kaneko:2018lattice} and inclusive decays~\cite{Hashimoto:2017wqo}. We use the M\"{o}bius domain wall action~\cite{Brower:2005qw} for all quarks, which has the advantage of including all relativistic effects for the heavy quarks, but necessitates extrapolating to physical $m_b$ from lower heavy quark masses, $m_h$. It is therefore important to assess the heavy quark scaling of the $B\to\pi\ell\nu$ form factors. 

Form factors for $B\to\pi\ell\nu$ can be related to vector matrix elements between $B$ and $\pi$ mesons by,
\begin{eqnarray}
  \langle \pi(k_\pi) | V^\mu | B(p_B) \rangle &=& f_+(q^2)\left[(p_B+k_\pi)^{\mu}-\frac{m_B^2-m_\pi^2}{q^2}q^\mu \right]+ f_0(q^2)\frac{m_B^2-m_\pi^2}{q^2}q^\mu,
\end{eqnarray}
where $q^\mu=p_B^\mu-k_\pi^\mu$, and $p_B$ and $k_\pi$ are the four-momenta of the $B$ meson and pion, respectively. It is also possible to express this in the language of Heavy Quark Effective Theory such that~\cite{Burdman:1993es},
\begin{eqnarray}
  \langle \pi(k_\pi) | V^{\mu} | B(p_B) \rangle &=&2\sqrt{m_B} \left[f_1\left(v\cdot k_\pi\right)v^\mu +f_2\left(v \cdot k_\pi\right)\frac{k^\mu}{v \cdot k_\pi}\right],
\end{eqnarray}
where $v^\mu=p^\mu_B/m_B$ and $E_\pi\equiv v \cdot k_\pi$. This parametrization of the form factors is convenient for lattice calculations since they relate to the spatial and temporal vector matrix elements by,
\begin{eqnarray}
  f_1(v\cdot k_\pi)+f_2(v\cdot k_\pi) &=& \frac{\langle \pi(k_\pi) | V^0 | B(p_B) \rangle}{2\sqrt{m_B}};\\
  f_2(v\cdot k_\pi) &=& \frac{\langle \pi(k_\pi)| V^i | B(v)\rangle}{2\sqrt{m_B}}\frac{v\cdot k_\pi}{k^i_\pi}, 
\end{eqnarray}
which in turn are related back to the vector and scalar form factors by,
\begin{eqnarray}
        f_+\left(q^2\right)&=&\sqrt{m_B}\left[\frac{f_2\left(v \cdot k\right)}{v\cdot k_\pi}+\frac{f_1\left(v\cdot k\right)}{m_B}\right];\\
        f_0\left(q^2\right)&=&\frac{2}{\sqrt{m_B}}\frac{m^2_B}{m^2_B-m^2_\pi}\left[{\frac{m^2_\pi}{\left(v\cdot k_\pi\right)^2}}\left(f_1\left(v\cdot k\right)+f_2\left(v \cdot k\right)\right)\right. \nonumber \\
          &&-\left.\frac{v\cdot k_\pi}{m_B}\left(f_1\left(v\cdot k\right)+\frac{m^2_\pi}{\left(v\cdot k_\pi\right)^2}f_2\left(v \cdot k\right) \right)\right].
\end{eqnarray}

One can then determine the $|V_{ub}|$ matrix element using branching fractions from experiment through,
\begin{eqnarray}
  \frac{\mathrm{d}\Gamma\left(B\rightarrow\pi\ell\nu\right)}{\mathrm{d}q^2}&=&\frac{G^2_F\left|V_{ub}\right|^2}{24\pi^3}\left|k_\pi\right|^3 \left|f_+(q^2)\right|^2.
\end{eqnarray}
  The scalar form factor $f_0(q^2)$ does not contribute in the limit of vanishing lepton masses, but it is nevertheless a useful and important quantity for lattice QCD calculations.

\section{Lattice Calculation}
We use $2+1$ flavour M\"{o}bius domain wall gauge ensembles that are tree-level Symanzik improved in our simulations. Lattice spacings included in this calculation are approximately $0.080~\mathrm{fm}$, $0.055~\mathrm{fm}$ and $0.044~\mathrm{fm}$, corresponding to $\beta=4.17$, $\beta=4.35$ and $\beta=4.47$, respectively. Pion masses range from $500~\mathrm{MeV}$ down to $\approx230~\mathrm{MeV}$, where we use a larger volume at $\beta=4.17$ for the lightest pion such that we maintain $m_\pi L >4$. Heavy quark masses are chosen to be $1.25^{2n}\times m_c$ for $n=0,1,2$, ensuring that $am_h<0.7$ to avoid large discretization effects from the heavy quark mass. This means that form factor results from our $D\to\pi\ell\nu$ study~\cite{Kaneko:2017xgg} are included as data points in this calculation. This gives us two values of the heavy quark mass on the coarsest ensemble and three values on the other two.

The $B$ meson is kept at rest in our calculations while we give pions momenta $|{\bf p}|^2=0,1,2,3$ in units of $(2\pi/L)^2$. We calculate correlators from all permutations of a given momentum and average these to improve our signal.


\section{Results}
For the form factor $f_1(v\cdot k_\pi)+f_2(v\cdot k_\pi)$ we use the global fit form,
\begin{eqnarray}
  f_1(v\cdot k_\pi)+f_2(v\cdot k_\pi) &=&  C_0\left(1+\sum^3_{n=1}C_{E^n_{\pi}}E^n_\pi\right)\left(1+C_{m^2_{\pi}}m^2_\pi\right)\left(1+C_{a^2}a^2\right)\left(1+\frac{C_{m_b}}{m_b}\right)\left(1+\chi_{\mathrm{log}}\right),\nonumber \\
  \label{eq:global_f1_f2}
\end{eqnarray}
and for the $f_2(v\cdot k_\pi)$ form factor we use,
\begin{eqnarray}
  f_2(v\cdot k_\pi)&=&\left[D_0\left(1+D_{E_\pi}E_\pi\right)\left(1+D_{m^2_\pi} m^2_\pi\right)\left(1+D_{a^2}a^2\right)\left(1+\frac{D_{m_b}}{m_b}\right)\left(1+\chi_{\mathrm{log}}\right)\right]\frac{E_\pi}{E_\pi+\Delta_{B}},\nonumber \\
  \label{eq:global_f2}
\end{eqnarray}
where $\chi_{\mathrm{log}}$ is a chiral log term and $\Delta_B$ is the $m_{B^*}$-$m_{B}$ splitting. We only use a term linear in the pion energy for the $f_2(v\cdot k_\pi)$ form factor, but include quadratic and cubic terms for $f_1(v\cdot k_\pi)+f_2(v\cdot k_\pi)$. A pole term appears in the $f_2(v\cdot k_\pi)$ fit form to account for the expected behaviour from pole dominance. 

\begin{figure}[!ht]
  \centerline{
    \includegraphics[width=0.61\textwidth]{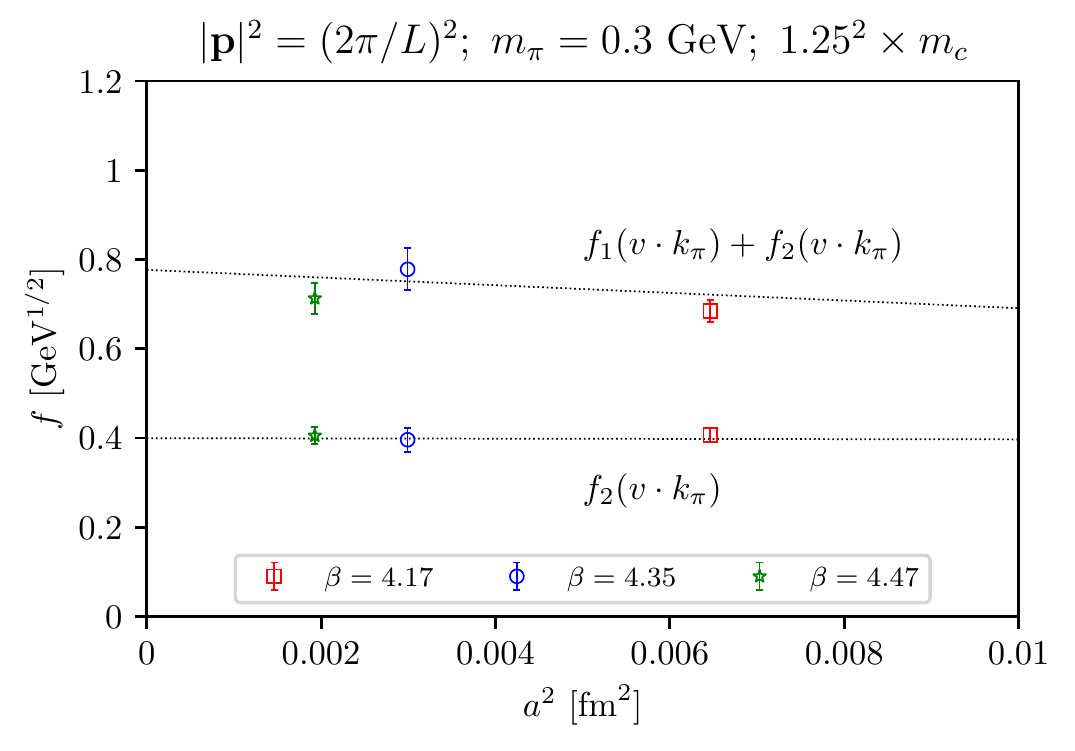}
  }
  \caption{Plot of the form factors $f_1(v\cdot k_\pi)+f_2(v\cdot k_\pi)$ and $f_2(v\cdot k_\pi)$ for $B\to\pi\ell\nu$ against $a^2$. The results shown are at $\beta=4.17$ (red squares), $\beta=4.35$ (blue circles) and $\beta=4.47$ (green stars), with the heavy quark masses fixed to $1.25^2\times am_c$ and $m_\pi\approx 300~\mathrm{MeV}$. The dotted black line shows the result of the global fit (set to the corresponding heavy quark and pion masses).}
  \label{fig:asq_dependence}
\end{figure}

In figure~\ref{fig:asq_dependence} we demonstrate the dependence on the lattice spacing by plotting the data points corresponding to $300~\mathrm{MeV}$ pions with a momentum of $|{\bf p}|^2=(2\pi/L)^2$, and $m_h=1.25^2\times m_c$ at each lattice spacing. The dotted lines are the corresponding results from the global fit.  We see only mild dependence on the lattice spacing for both form factors, and our fit form describes the data well.

\begin{figure}[!ht]
  \centerline{
    \includegraphics[width=0.61\textwidth]{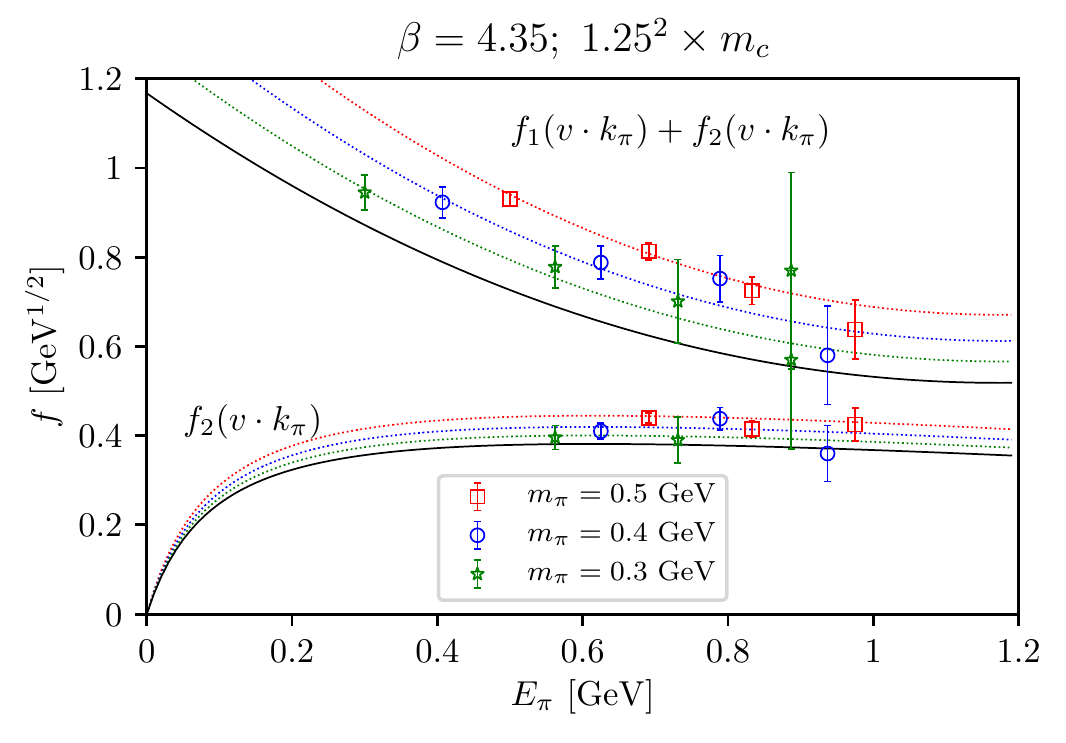}
  }
  \caption{Plot of the $m_\pi$ dependence of form factors for $B\to\pi\ell\nu$ against pion energy. The points and the corresponding dotted lines denote results for $m_\pi$ at $500~\mathrm{MeV}$ (red squares), $400~\mathrm{MeV}$ (blue circles) and $300~\mathrm{MeV}$ (green stars) from the global fit fixed to heavy quark masses $1.25^2\times am_c$ on the $\beta=4.35$ ensemble. The solid black line shows the extrapolation to the physical value of $m_\pi=135~\mathrm{MeV}$.}
  \label{fig:mpi_dependence}
\end{figure}

The pion mass dependence is shown in figure~\ref{fig:mpi_dependence}, where we plot the form factors against the pion energy for three values of the pion mass ($500~\mathrm{MeV}$, $400~\mathrm{MeV}$ and $300~\mathrm{MeV}$). The black solid line is the result of an extrapolation to the physical pion mass, $m_\pi=135~\mathrm{MeV}$. Again, the data is well-described by our chosen fit form. The dependence on $m_\pi$ is clearly milder for the $f_2(v\cdot k_\pi)$ form factor.

\begin{figure}[!ht]
  \centerline{
    \includegraphics[width=0.61\textwidth]{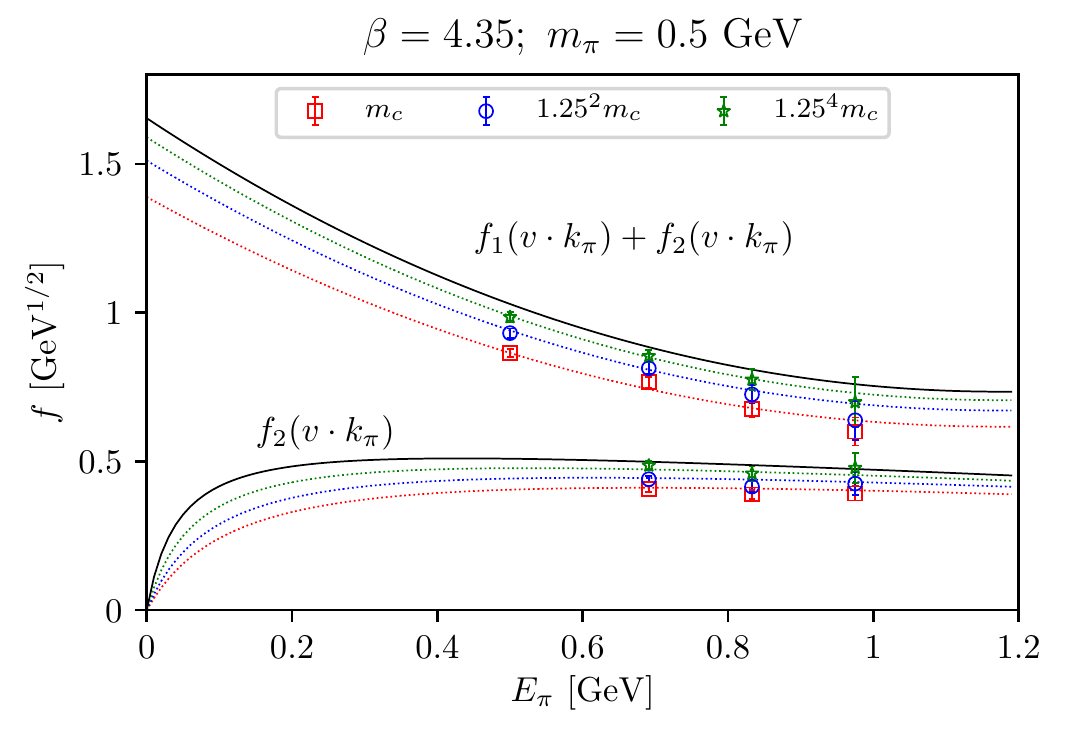}
  }
  \caption{Plot of the heavy quark dependence of form factors for $B\to\pi\ell\nu$ against pion energy. The red squares, blue circles and green stars -- and the corresponding dotted lines -- denote results from the global fit fixed to heavy quark masses $am_c$, $1.25^2\times am_c$ and $1.25^4\times am_c$, respectively, on the $\beta=4.35$ ensemble with $m_\pi\approx 500~\mathrm{MeV}$. The solid black line shows the extrapolation to the physical value of $m_b$.}
  \label{fig:heavy_mass_dependence}
\end{figure}

Finally, we make a similar plot for the dependence on the heavy quark mass in figure~\ref{fig:heavy_mass_dependence}. We show results from the $\beta=4.35$ ensemble with $m_\pi=500~\mathrm{MeV}$. The solid black line is the result from the global fit set to the physical value of $m_b$.  We only include a term of order $1/m_b$ in our global fit forms, and conclude from this plot that this is sufficient.

\begin{figure}[!ht]
  \centerline{
    \includegraphics[width=0.61\textwidth]{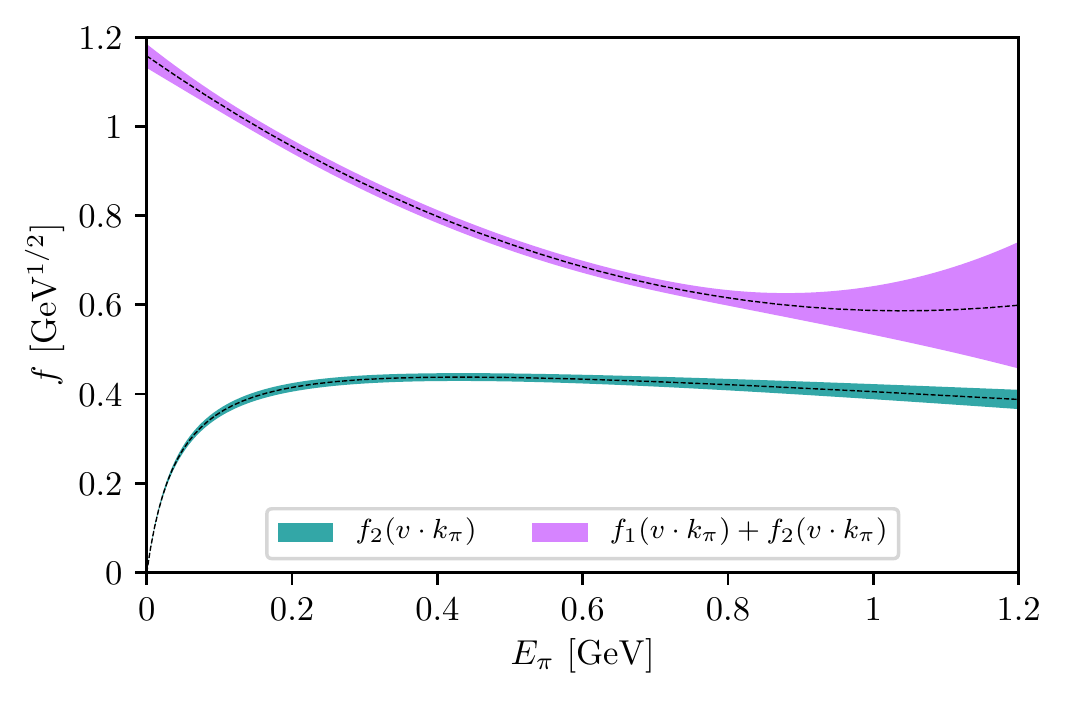}
  }
  \caption{Results for form factors $f_1(v\cdot k_\pi)+f_2(v\cdot k_\pi)$ and $f_2(v\cdot k_\pi)$ from the global fit forms in eq.~(\ref{eq:global_f1_f2}) and eq.~(\ref{eq:global_f2}), respectively. The dashed black lines show the central values and the coloured bands depict the statistical uncertainty only.}
  \label{fig:global_fit}
\end{figure}

The overall results from these fits to our data is shown in figure~\ref{fig:global_fit} where the upper band is the result for $f_1(v\cdot k_\pi)+f_2(v\cdot k_\pi)$ and the lower for $f_2(v\cdot k_\pi)$. The form factors are shown after extrapolations to physical $m_\pi$ and $m_b$, and to the continuum limit. We find $\chi^2/\mathrm{dof}=0.88$ for $f_1(v\cdot k_\pi)+f_2(v\cdot k_\pi)$ and $\chi^2/\mathrm{dof}=0.69$ for $f_2(v\cdot k_\pi)$.

\subsection{$z$-expansion}

Experimental branching fraction results are available for $B\to\pi\ell\nu$ from the Belle and BaBar collaborations~\cite{delAmoSanchez:2010af,Ha:2010rf,Lees:2012vv,Sibidanov:2013rkk}. In contrast to the lattice, this is most precise at low $q^2$ and thus a determination of $|V_{ub}|$ benefits from a robust extrapolation of the lattice data towards this region. This is typically done through the use of a so-called $z$-parameter expansion where we have used the form,
\begin{equation}
  f_{+/0}=\frac{1}{P(q^2)}\sum_{n=0} b^{(n)}z^n.
  \label{eq:zexpansion}
\end{equation}
The momentum transfer is now mapped onto $z$ through,
\begin{equation}
  z=\frac{\sqrt{t_+-q^2}-\sqrt{t_+-t_0}}{\sqrt{t_+-q^2}+\sqrt{t_+-t_0}},
\end{equation}
where $t_+=(m_B+m_{\pi})^2$. We have chosen $t_0=0$ so that $z=0$ corresponds to $q^2=0$. For the pole factor $P(q^2)$ in eq.~(\ref{eq:zexpansion}) we use $1$ for $f_0(q^2)$ and $(1-q^2/M_{B^*})$ for $f_+(q^2)$.

\begin{figure}[!ht]
  \centerline{
    \includegraphics[width=0.61\textwidth]{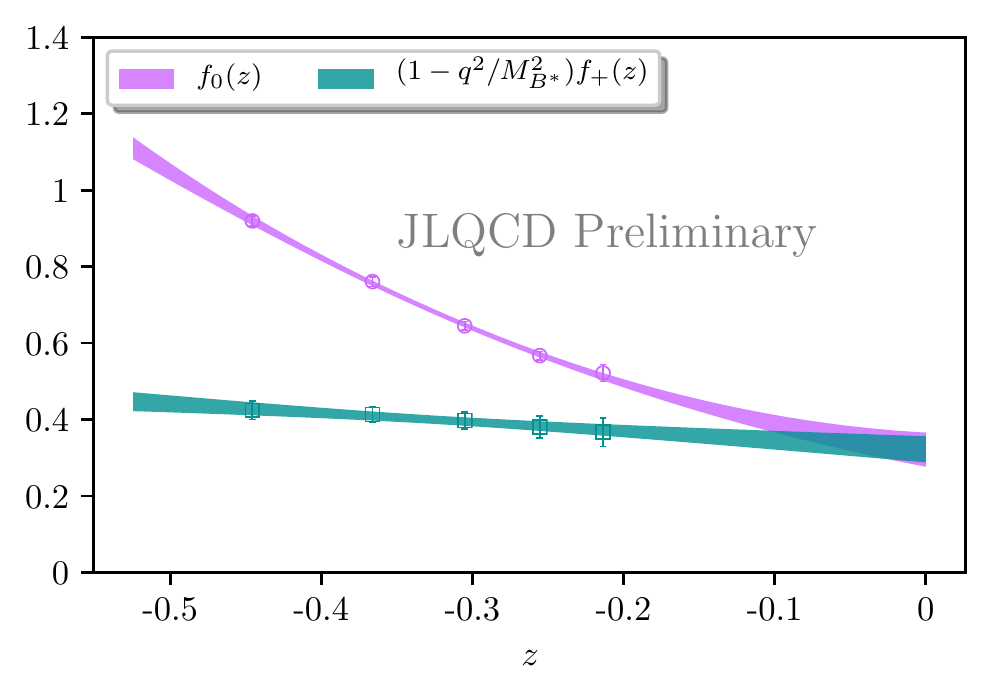}
  }
  \caption{$z$-parameter expansion of the two form factors across the entire $q^2$ range. Synthetic data points from the global fit results are shown. The bands only depict statistical uncertainties.}
  \label{fig:zexpansion}
\end{figure}

\begin{figure}[!ht]
  \centerline{
    \includegraphics[width=0.61\textwidth]{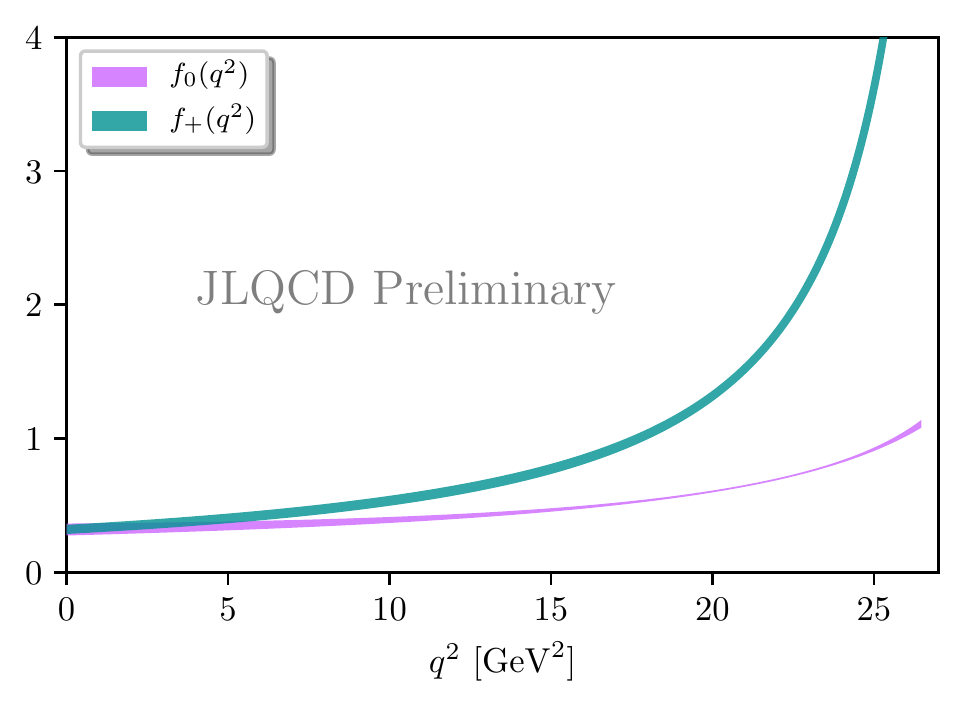}
  }
  \caption{Form factors across the entire $q^2$ range after a $z$-parameter expansion fit using synthetic data from the global fit results. The bands only depict statistical uncertainties.}
  \label{fig:form_factors}
\end{figure}

A fit result for this expansion is shown in figure~\ref{fig:zexpansion}, where we have used synthetic data from the results of our global fits in the physical $m_\pi$, $m_b$ and continuum limits. We convert back to $q^2$ and again plot the form factors across the entire momentum-transfer range in figure~\ref{fig:form_factors}. The bands only show statistical uncertainties for each of the form factors.

\section{Outlook}
We have presented an update of our work on $B\to\pi\ell\nu$ form factors using M\"{o}bius domain wall fermions. We find the dependences on lattice spacing and pion mass to be well under control. Similarly, the scaling of the heavy quark mass is well-described by a $1/m_b$ HQET expansion. By using a $z$-parameter expansion on synthetic data after extrapolations to physical $m_\pi$, $m_b$ and the continuum limit, we obtain preliminary results for the form factors $f_+(q^2)$ and $f_0(q^2)$ across the entire $q^2$ range with statistical uncertainties only.

We now work towards a complete error budget for these form factors, after which we will fit our results alongside differential branching fractions from experiment to make a new determination of the CKM matrix element $|V_{ub}|$.

\acknowledgments
Numerical simulations were performed on the Blue Gene/Q supercomputer under its Large Scale Simulation Program (No. 16/17-14) and on JCAHPC's Oaksforest-PACS. This work is supported in part by JSPS KAKENHI Grant Number 18H03710, by MEXT as ``Priority Issue on post-K computer'' (Elucidation of the Fundamental Laws and Evolution of the Universe) and the Joint Institute for Computational Fundamental Science (JICFuS).
\bibliographystyle{JHEP}
\bibliography{sources}

\providecommand{\href}[2]{#2}\begingroup\raggedright\begin{thebibliography}{1}

\bibitem{Kaneko:2018lattice}
T.~Kaneko, Y.~Aoki, B.~Colquhoun, S.~Hashimoto and H.~Fukaya, \emph{{$B\to
  D^{(*)}\ell\nu$ form factors from $N_f=2+1$ QCD with Moebius
  domain-wall-quarks}}, {\emph{\pos{PoS(LATTICE2018)311}} (2018) }.

\bibitem{Hashimoto:2017wqo}
S.~Hashimoto, \emph{{Inclusive semi-leptonic B meson decay structure functions
  from lattice QCD}}, \href{https://doi.org/10.1093/ptep/ptx052}{\emph{PTEP}
  {\bfseries 2017} (2017) 053B03}
  [\href{https://arxiv.org/abs/1703.01881}{{\ttfamily 1703.01881}}].

\bibitem{Brower:2005qw}
R.~C. Brower, H.~Neff and K.~Orginos, \emph{{Mobius fermions}},
  \href{https://doi.org/10.1016/j.nuclphysbps.2006.01.047}{\emph{Nucl. Phys.
  Proc. Suppl.} {\bfseries 153} (2006) 191}
  [\href{https://arxiv.org/abs/hep-lat/0511031}{{\ttfamily hep-lat/0511031}}].

\bibitem{Burdman:1993es}
G.~Burdman, Z.~Ligeti, M.~Neubert and Y.~Nir, \emph{{The Decay $B
  \to\pi\ell\nu$ in heavy quark effective theory}},
  \href{https://doi.org/10.1103/PhysRevD.49.2331}{\emph{Phys. Rev.} {\bfseries
  D49} (1994) 2331} [\href{https://arxiv.org/abs/hep-ph/9309272}{{\ttfamily
  hep-ph/9309272}}].

\bibitem{Kaneko:2017xgg}
{\scshape JLQCD} collaboration, T.~Kaneko, B.~Colquhoun, H.~Fukaya and
  S.~Hashimoto, \emph{{D meson semileptonic form factors in $N_f$ = 3 QCD with
  M\"{o}bius domain-wall quarks}},
  \href{https://doi.org/10.1051/epjconf/201817513007}{\emph{EPJ Web Conf.}
  {\bfseries 175} (2018) 13007}
  [\href{https://arxiv.org/abs/1711.11235}{{\ttfamily 1711.11235}}].

\bibitem{delAmoSanchez:2010af}
{\scshape BaBar} collaboration, P.~del Amo~Sanchez et~al., \emph{{Study of $B
  \to \pi \ell \nu$ and $B \to \rho \ell \nu$ Decays and Determination of
  $|V_{ub}|$}}, \href{https://doi.org/10.1103/PhysRevD.83.032007}{\emph{Phys.
  Rev.} {\bfseries D83} (2011) 032007}
  [\href{https://arxiv.org/abs/1005.3288}{{\ttfamily 1005.3288}}].

\bibitem{Ha:2010rf}
{\scshape Belle} collaboration, H.~Ha et~al., \emph{{Measurement of the decay
  $B^0\to\pi^-\ell^+\nu$ and determination of $|V_{ub}|$}},
  \href{https://doi.org/10.1103/PhysRevD.83.071101}{\emph{Phys. Rev.}
  {\bfseries D83} (2011) 071101}
  [\href{https://arxiv.org/abs/1012.0090}{{\ttfamily 1012.0090}}].

\bibitem{Lees:2012vv}
{\scshape BaBar} collaboration, J.~P. Lees et~al., \emph{{Branching fraction
  and form-factor shape measurements of exclusive charmless semileptonic B
  decays, and determination of $|V_{ub}|$}},
  \href{https://doi.org/10.1103/PhysRevD.86.092004}{\emph{Phys. Rev.}
  {\bfseries D86} (2012) 092004}
  [\href{https://arxiv.org/abs/1208.1253}{{\ttfamily 1208.1253}}].

\bibitem{Sibidanov:2013rkk}
{\scshape Belle} collaboration, A.~Sibidanov et~al., \emph{{Study of Exclusive
  $B \to X_u \ell \nu$ Decays and Extraction of $\|V_{ub}\|$ using Full
  Reconstruction Tagging at the Belle Experiment}},
  \href{https://doi.org/10.1103/PhysRevD.88.032005}{\emph{Phys. Rev.}
  {\bfseries D88} (2013) 032005}
  [\href{https://arxiv.org/abs/1306.2781}{{\ttfamily 1306.2781}}].

\end{thebibliography}\endgroup

\end{document}